\documentclass{sig-alternate}

\usepackage{fixltx2e}
\usepackage{graphicx}
\usepackage{color}
\usepackage{rotating}
\usepackage{amsfonts}
\usepackage{pifont}
\usepackage{program} 
\usepackage{float}

\begin{document}

\title{Connecting Mobile Things to Global Sensor Network Middleware using System-generated Wrappers}

\author{Charith Perera\textsuperscript{*+} \and Arkady Zaslavsky\textsuperscript{+}  \and Peter Christen\textsuperscript{*}   \and Ali Salehi\textsuperscript{+} \and Dimitrios Georgakopoulos\textsuperscript{+}\\
\\ 
\textsuperscript{*}Research School of Computer Science, The Australian National University, \\Canberra, ACT 0200, Australia
\\   
\textsuperscript{+}CSIRO ICT Center, Canberra, ACT, 2601, Australia      
}


%

\maketitle

\begin{abstract}

Internet of Things (IoT) will create a cyberphysical world where all the things around us are connected to the Internet, sense and produce ``big data'' that has to be stored, processed and communicated with minimum human intervention. With the ever increasing emergence of new sensors, interfaces and mobile devices, the grand challenge is to keep up with this race in developing software drivers and wrappers for IoT things. In this paper, we examine the approaches that automate the process of developing middleware drivers/wrappers for the IoT things. We propose ASCM4GSN architecture to address this challenge efficiently and effectively. We demonstrate the proposed approach using Global Sensor Network (GSN) middleware which exemplifies a cluster of data streaming engines. The ASCM4GSN architecture significantly speeds up the wrapper development and sensor configuration process as demonstrated for Android mobile phone based sensors as well as for Sun SPOT sensors.

\end{abstract}

\keywords{Internet of Things, Sensor Networks, Global Sensor Network Middleware, Mobile Sensors}
%
%
%

%

\category{C.2.1}{Computer-Communication Networks}{Network Architecture and Design}[Sensor networks]
\category{C.2.4}{Computer-Communication Networks}{Distributed Systems}
\category{D.2.11}{Software Engineering}{Software Architectures}
\category{H.3.4}{Information Storage and Retrieval}{Systems and Software}[Current awareness systems]


\section{Introduction}
\label{sec:Introduction}


The term Internet of Things (IoT) was firstly coined by Kevin Ashton \cite{P065} in a presentation in 1998. Further expanding this idea, the European Union has defined the above vision as ``\textit{The IoT allows people and things to be connected Anytime, Anyplace, with Anything and Anyone, ideally using Any network and Any service} \cite{P019}''. It is expected that 50 to 100 billion devices will be connected to the Internet by 2020. According to the BCC Research \cite{P255}, global market for sensors was around \$56.3 billion in 2010. In 2011, it was around \$62.8 billion. Global market for sensors is expected to increase up to \$91.5 billion by 2016, at a compound annual growth rate  of 7.8\%. The connection and configuration of these sensor devices are not feasible to be done manually. Automation is essential to achieve the vision of IoT. This is the challenge we have addressed.

There is an increasing trend of developing middleware solutions in order to connect sensors and actuators to the Internet. These middleware solutions support fast and simple deployment of sensor networks. GSN \cite{P022}, Sgroi et al. \cite{P112}, Hourglass \cite{P033}, HiFi \cite{P111}, IrisNet \cite{P089}, and EdgeServers \cite{P113} are some of the major middleware solutions. These systems share a common objective with minor differences in features and functionality. The GSN solution provides advanced and sophisticated functionality. Therefore, we decided to use GSN as the sensor network middleware to exemplify our proposed solution.

One of the major drawback in these existing middleware solutions is connectivity and configurability. Sensors come with APIs that provide software interfaces to retrieve sensor data to the middleware solutions or applications. Different middleware solutions use different mechanisms to retrieve data from sensors. The solutions are referred using different terms such as \textit{wrappers}, \textit{gateways}, \textit{handlers}, \textit{proxies}, \textit{mediators}, etc. For example, GSN has a concept called \textit{Wrapper} \cite{P022}. Each sensor should have a supported wrapper to be attached to the GSN server, in order to communicate with the sensor hardware. These wrappers need to be developed manually. This is an overwhelming task. There are many sensor devices and smart objects \cite{P041} that come to the market regularly. Therefore, developing wrappers or similar solutions manually is not a scalable and feasible approach. Thus, we investigate the methods that can significantly automate this process.

The rest of the paper is organised as follows. Section \ref{sec:Sensor Networks} presents an overview on sensor networks. Section \ref{sec:Global Sensor Network} describes Global Sensor Network (GSN) middleware in brief. GSN wrapper is briefly discussed in Section \ref{sec:Gsn Wrapper}. The life cycle of the wrapper is presented in Section \ref{sec:Gsn Wrapper'S Life Cycle}. Section \ref{sec:The Problem} explains the problem that we have addressed in detail. Our proposed solution is presented in detail in Section \ref{sec:Our Approach}. The Sensor Device Definition (SDD) files are explained in Section \ref{sec:Sensor Device Definition}. Section \ref{sec:Implementation} presents the experiment of connecting Android mobile devices to GSN. Finally, Section \ref{sec:Related Work} presents the related work and is followed by a conclusion.

\begin{figure*}
 \centering
 \includegraphics[scale=.75]{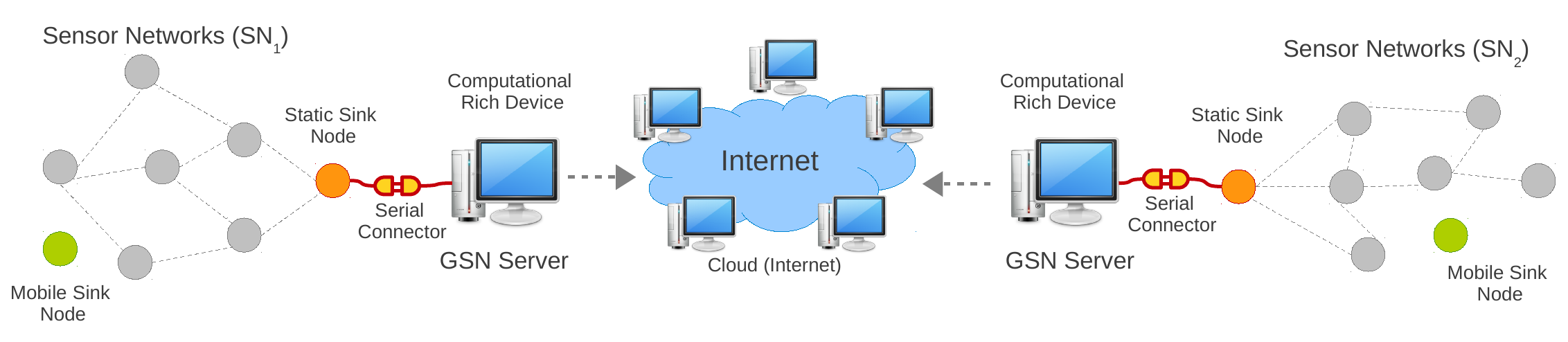}
\vspace{-0.7cm}
 \caption{Sensor Networks Communicating Over Internet}
 \label{Sensor Networks Communicating Over Internet}
\end{figure*}


\section{Sensor Networks}
\label{sec:Sensor Networks}

Sensor networks are the major enabler of the IoT. A \textit{sensor network} \cite{P009} comprises one or more sensor nodes which communicate between each other using wired and wireless means. Each sensor node has the capability to sense, communicate and process data. In sensor networks, sensors can be homogeneous or heterogeneous.  These sensors are deployed in densely manner around the phenomenon which we want to sense \cite{P009}. These sensor nodes are typically low-cost and small in size that enable large deployments. 

Today, increasing number of mobile sensors becoming available in the market as well as in sensor network deployments. Mobile sensors are capable of sensing while moving from one location to another. They add more efficiency and accuracy to the sensor networks, because a single mobile sensor can sense a larger area than a static sensor. Sensors built-in to the mobile phones can behave as mobile sensors and make a significant contribution in community sensing. For an experiment, we connected a mobile phone to the GSN to collect data via  built-in sensors using manual wrapper development approach. We were able to identify the difficulties and challenges related to manual wrapper development approach as presented in Section \ref{sec:Implementation}. We proposed our ASCM4GSN approach to mitigate these difficulties.

Before IoT, sensor networks have been used in domains such as health, military and home \cite{P009}. Today, sensor networks and IoT share substantial amount of similarities. However, most of the sensor network researches have been focused on low level design and deployment issues. For example, fault tolerance, scalability, production cost, hardware constraints, sensor network topology, environment, transmission media, unpredictability, heterogeneity, energy efficient counting, localisation algorithms and power consumption are some of the leading research areas in the sensor networks \cite{P009, P010, P037, P051}. High level research areas such as data fusion and processing, have gained significant attention only in recent years. 

Sensor network deployment has been considered as a difficult task in early days due to the heterogeneity of sensors. Developers and engineers had to deal with low level programming tasks in order to connect these sensor nodes together and get the sensor data into applications. A number of frameworks and middleware solutions have been developed in order to make this process easier. Global Sensor Network (GSN) \cite{P050} is such a middleware solution that enables zero-configuration deployment. It is used widely in over ten EU/Swiss funded research projects \cite{P167}. Figure \ref{Sensor Networks Communicating Over Internet} summarises our discussion and shows how the components we discussed above fit in real world. The next section introduces the Global Sensor Network (GSN) middleware.


\section{Global Sensor Network}
\label{sec:Global Sensor Network}

The Global Sensor Network (GSN) \cite{P022, P050} is a platform aimed at providing flexible middleware to address the challenges of sensor data acquisition, integration and distributed query processing. It is a generic data stream processing engine. GSN has gone beyond the traditional sensor network research efforts such as routing, data aggregation, and energy optimisation. The design of GSN is based on four basic principles: simplicity, adaptivity, scalability, and light-weight implementation. GSN simplifies the process of connecting heterogeneous sensor devices to applications. Specifically, GSN provides the capability to integrate, discover, combine, query, and filter sensor data through a declarative XML-based language and enables zero-programming deployment and management. The above reasons lead us to choose GSN as our sensor network middleware over other alternative solutions.

The GSN adopts container based architecture. A detailed explanation is provided in \cite{P022}. \textit{Virtual Sensor} is the key element in the GSN. A virtual sensor can be any kind of data producer, for example, a real sensor, a wireless camera, a desktop computer, a mobile phone, or any combination of virtual sensors. Typically, a virtual sensor can have multiple input data streams but have only one output data stream.



A \textit{Wrapper} is a piece of code that does the data acquisition from a specific type of sensor device. The GSN is capable of retrieving data from various data sources. Wrappers transform the raw data into the GSN standard data model that can be queried and manipulated later. All the wrapper classes need to extend the \textit{AbstractWrapper} class. Typically, third party libraries are initialised in the wrapper constructor. Each sensor needs to have a specific wrapper that can be used to retrieve raw sensor data. In order to connect a Mica2 \cite{P152} sensor, for example, the GSN should have a corresponding wrapper that can talk to Mica2 sensor and retrieve data from it. Currently, the GSN provides wrappers for all TinyOS \cite{P149} based sensors, RFID sensors, web cams, actuators, etc. Likewise, in order to connect an Android phone's built-in sensors to the GSN, it has to have a wrapper that can retrieve raw sensor data from Android phones. We discuss GSN wrappers in general and wrapper's life cycle in details in the next sections. 


\section{Gsn Wrapper}
\label{sec:Gsn Wrapper}

\begin{figure}[H]
 \centering
 \includegraphics[scale=.9]{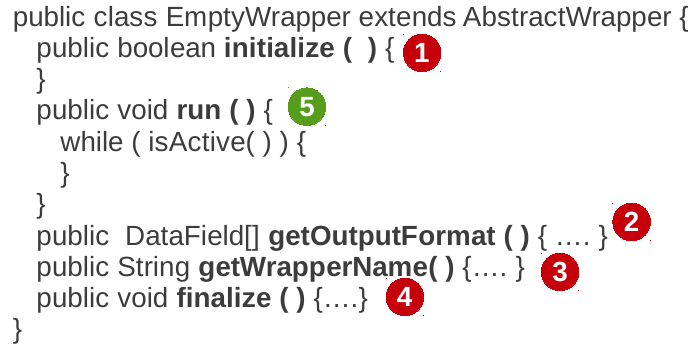}
 \caption{GSN Wrapper}
 \label{GSN Wrapper}	
\end{figure}

In this section, we discuss GSN wrappers. As we explained earlier, each and every sensor that needs to be connected to GSN should have a corresponding wrapper. The Figure \ref{GSN Wrapper} depicts a basic code structure for a GSN wrapper.

All the wrappers need to extend the Java class \newline \textit{gsn.wrapper.AbstractWrapper}. Therefore, all the wrappers are subclasses of \textit{AbstractWrapper}. There are four methods that need to be implemented by the subclasses. Those methods are numbered 1-4 in the Figure \ref{GSN Wrapper}. The methods are 1. \textit{boolean initialise()}, 2. \textit{void finalise()}, 3. \textit{String getWrapperName()}, and 4. \textit{DataField[] getOutputFormat()}.


A new thread is created for each wrapper in GSN. After creating the wrapper object, \textit{initialise()} method is called as shown as (1) in Figure \ref{GSN Wrapper}. All the communication using third party libraries should happen within this method. For example, a camera wrapper may talk to a third party API in order to talk to the camera and retrieve the camera images. In an \textit{AndroidWrapper}, \textit{initialise()} method creates a socket and waits until the client mobile phone sends the data packets. 

The method \textit{finalise()} is called at the end of the wrapper's life cycle. This method can be used to close all the connections that established with the outer world by the wrapper. Concretely, all the resources acquired during the \textit{initialise()} method should be released here. For example, in the Android wrapper, all the client communication resources such as sockets and ports are released in this method.

The method \textit{getWrapperName()} returns the name of the wrapper. The method \textit{getOutputFormat()} returns a \textit{DataField} object that provides a description of the data structure produced by the wrapper. The \textit{run()} method is responsible for retrieving sensor data from sensors and transform them into GSN data model. All the sensor specific API calls need to be done inside this method. This is the most complex section of the class. The content of this method is hard to generate automatically due to third party library dependencies. The auto generation of GSN wrappers is discussed in Section \ref{sec:Our Approach}.


\section{Gsn Wrapper'S Life Cycle}
\label{sec:Gsn Wrapper'S Life Cycle}

The life cycle of a wrapper begins with the initiation of Virtual Sensor Definition (VSD) \cite{P167} file. When a user defines a VSD file, it triggers the virtual sensor creation process. This process triggers the specified wrapper to be generated. 

\begin{figure}[!h]
 \centering
 \includegraphics[scale=1]{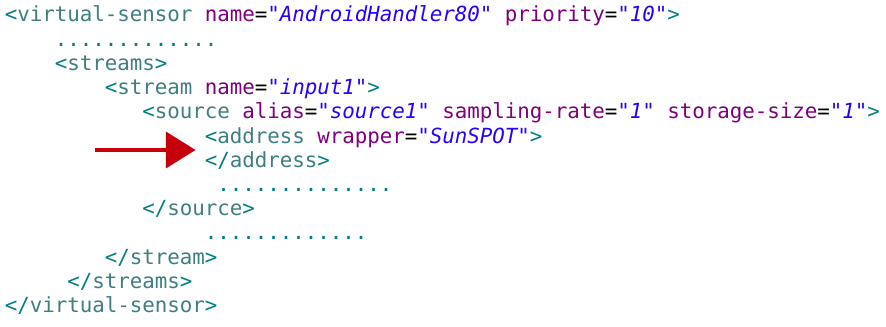}
 \caption{Virtual Sensor}
 \label{Virtual Sensor}	
\end{figure}

The wrapper that corresponds to each stream source (i.e. sensor) is defined under the address element in the VSD file. For example, the VSD file segment depicted in Figure \ref{Virtual Sensor} triggers the SunSPOT \cite{P382} wrapper to be instantiated. Virtual sensor creation process sends a Wrapper Connection Request (WCR) to the wrapper repository \cite{P167} in the GSN server. A Wrapper Connection Request is an object which contains a wrapper name and its initialisation parameters as defined in the \textit{Virtual Sensor}. Sequentially, the following steps are followed:

 \begin{figure}[H]
 \centering
 \includegraphics[scale=1]{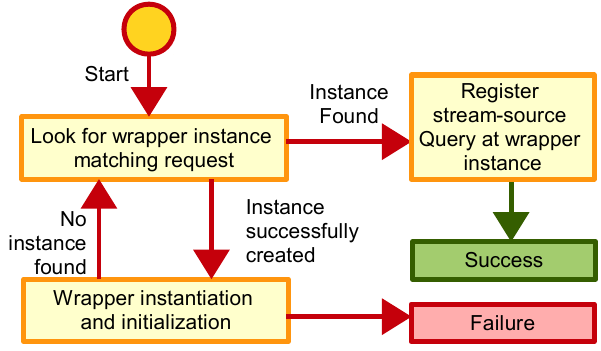}
 \caption{Wrapper Life Cycle}
 \label{Wrapper Life Cycle}	
\end{figure}

%
%

First, wrapper repository looks for a wrapper instance that matches to the WCR. If found, then the stream-source query will be registered with the wrapper and returns true.

If there isn't any wrapper object that matches WCR in the repository, the wrapper repository generates a new appropriate wrapper object. Then, the newly created object would be added to the wrapper repository. Finally, the stream-source query will be registered with the wrapper and returns true.

If there isn't any wrapper object that matches WCR in the repository and wrapper repository does not have an appropriate wrapper class to be instantiate, then returns false. The virtual sensor loader fails to load a virtual sensor if at least one of the stream sources required by an input stream fails. For example, if user defines a virtual sensor as depicted in the Figure \ref{Virtual Sensor} and if the GSN wrapper repository does not have a SunSPOT wrapper, then the virtual sensor would fail. Figure \ref{Wrapper Life Cycle} summaries the life cycle of the GSN wrappers. We proposed to extend this process in our solution. The details are explained in Section \ref{sec:Our Approach}.


\section{The Challenge of Improving Efficiency in Connecting Things}
\label{sec:The Problem}

\begin{figure}[h]
 \centering
 \includegraphics[scale=.8]{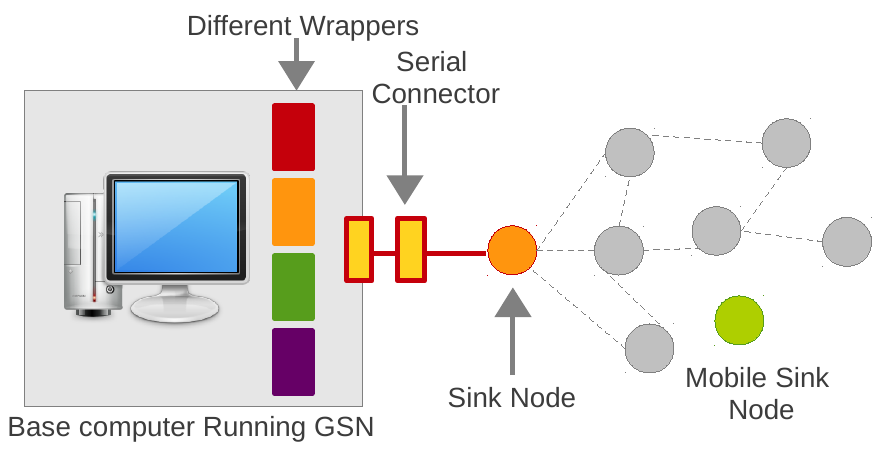}
 \caption{Current GSN Data Acquisition Architecture}
 \label{Current GSN Data Retrieve Method}	
\end{figure}

We introduced the problem in brief in earlier sections. Let's discuss it in details. Almost all the sensors come with third-party libraries or API released by the sensor manufacturer. If we want to retrieve sensor readings, we need to access the sensor hardware through these provided third party libraries. This stays true when we want to develop sensor networks using sensor network middleware solutions. For example, sensor network middleware solution such as GSN provides features such as window based continuous sensor data querying. In order to accomplish this task, GSN should retrieve sensor data from the sensor devices and organise them according to GSN specific data model. Most of the sensor network middleware solutions are good at providing high-level features such as querying which deal with internal data structure. However, the biggest problem is getting the sensor data from the sensor hardware devices into middleware solutions. This challenge has been addressed by different sensor network middleware solutions using different mechanisms. For example, GSN uses \textit{Wrappers} to accomplish this task. The Figure \ref{Current GSN Data Retrieve Method} shows the current GSN Data Acquisition Architecture.

However, we identify two drawbacks in current GSN data acquisition architecture. First problem is that these wrappers need to be developed manually by the programmers. For example, if we want to connect a SunSPOT sensor into the GSN, developers have to develop a wrapper that is specific for SunSPOT. This wrapper will use the libraries provided by the SunSPOT manufacturer to talk to the SunSPOT sensor and retrieve sensor data. When the sensor devices get updated, GSN developers have to update these wrappers as well. That means developers have to keep on updating these wrappers. This process decreases the scalability and it also requires more effort and cost.

The second problem is lack of code sharing. For example, one project may develop a wrapper for SunSPOT sensors. However, there is no mechanism to distribute this wrapper among other projects. The same wrapper may be developed by different project groups. It reduces the effectiveness and efficiency due to repetition. We have addressed these two issues, first by automating wrapper generation and second by developing a cloud repository. We discuss the details of our proposed solution in Section \ref{sec:Our Approach}.

%

Let us discuss some of the possible approaches that we can follow to solve the problem presented above. By evaluating several sensor devices and IoT middleware systems, it was understood that the method (steps) of connecting a device to an IoT middleware system is significantly similar. The most commons steps can be explained as bellow.

\begin{itemize}
\setlength\itemsep{-0.1ex}

 \item \textit{\textbf{Acquire Manufacturers' APIs}}: As we mentioned earlier, each sensor device comes with an API. In some instances, APIs are common across all the sensors devices produce by the manufacturer. In other cases, APIs are strictly related to specific type of sensors. However, middleware system should use these APIs to communicate with the sensor devices. Therefore, the first step is to identify the path to the APIs.

 \item \textit{\textbf{Acquire System Configuration Details}}: Most of the IoT middleware solutions need system configuration details such as IP addresses, ports and protocols to communicate with the sensor devices. Sometimes, it is possible to identify these details automatically and otherwise users may need to enter them manually. Therefore, identifying the machine specific and platform specific information is also a critical step.

 \item \textit{\textbf{Initiate the Data Structure}}: Every middleware maintains its own data structure. Once the sensor devices are connected to the middleware, the data sensed by the sensors need to be stored in these data structures. Therefore, identify the required data structure and allocate them is a major steps.

 \item \textit{\textbf{Initiate the Communication between IoT Middleware and Sensor Device}}: The communication between a sensor device and a IoT middleware solution usually starts by initiating the communication. The exact process could be varied among different sensor platforms. This initiation does not exchange any sensor data, but it opens and establishes the necessary ports and paths to proceed. This step also allocates the required resources and makes them ready for the data communication. This step further ensures that both sensor and middleware is ready to communicate between each other.

\item \textit{\textbf{Data Communication}}: This is the most important step. Based on the communication path established in the previous step, the sensor and the middleware will communicate either in push or pull method. As a result, middleware will receive sensor data periodically. The IoT middleware can store the sensor data on the data structure which prepared in an earlier step.

\item \textit{\textbf{Close the Communication and Release the Resources}}: All the connections established between the sensor device and the middleware system needs to be closed. Furthermore, the relate resources need to be released, so they are available for used by other operations.

\end{itemize}

We encapsulated these steps into five segments in the Sensor Device Definition (SDD) files as discussed in Section \ref{sec:Sensor Device Definition}. After identifying these commonalities, we investigate the methods of simplifying the process of connecting the sensors to IoT middleware systems.

Every IoT middleware has its own way of communicating with sensor devices. Mostly, there are dedicated handlers to accomplish this task. For example, in GSN, the handlers are called \textit{wrappers} as discussed in Section \ref{sec:Gsn Wrapper}. In other approaches, these handlers are called \textit{gateways}, \textit{proxies}, \textit{mediators}, etc. Furthermore, different technologies can also be employed to develop these handlers such as web services, RESTful APIs, native code, etc. From the previous work conducted by different researchers, it has been identified that native code gives better performance in term of scalability and efficiency compared to other technologies such as web services \cite{P212}. Therefore, we decided to use native code to develop the handler, in our case GSN's \textit{wrappers}. 

\begin{figure}[h]
 \centering
 \includegraphics[scale=1.2]{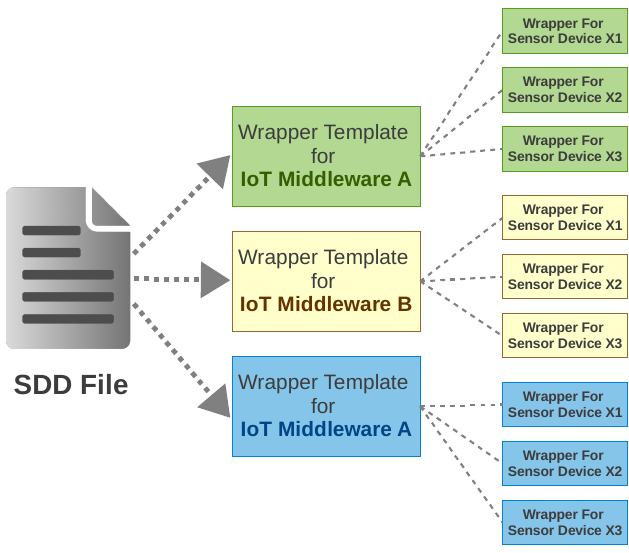}
 \caption{The Wrapper Generation Process}
 \label{The Wrapper Generation Process}
\vspace{-0.50cm}	
\end{figure}

The process of automated wrapper generation is depicted in Figure \ref{The Wrapper Generation Process}. Once we define a SDD file for a specific sensor as explained in Section \ref{sec:Sensor Device Definition}, it can be used to develop wrappers for different IoT middleware systems. As we mentioned earlier, every IoT middleware has its own component similar to wrappers. We can combine the wrapper template of an IoT middleware and a SDD file to generate a Middleware specific wrapper.  Wrapper template explains the basic structure of a wrapper such as functions, methods, data structures, etc.

\section{ASCM4GSN Architecture}
\label{sec:Our Approach}

The two main problems in the current approach are addressed as follows. We propose Automated Sensor Configuration Model For Global Sensor Network (ASCM4GSN) architecture to address these issues. We introduce a \textit{Automated Wrapper Generation Layer}. As the name implies, this layer automate the process of wrapper generation. The process can be explained as follows.

As we discussed in Section \ref{sec:Gsn Wrapper'S Life Cycle}, the wrapper generation always begins with a Virtual Sensor Definition (VSD). When a VSD mentions a name of a wrapper that GSN currently does not have in the wrapper repository, first, GSN searches the Sensor Device Definition Local Repository (SDDLR) to look for a matching Sensor Device Definition (SDD) file. We discuss SDD file in detail in the next Section. For now, SDD file can be explained as a specification file that contains all the information that is required to generate a wrapper class.

\begin{figure}[h]
 \centering
 \includegraphics[scale=.9]{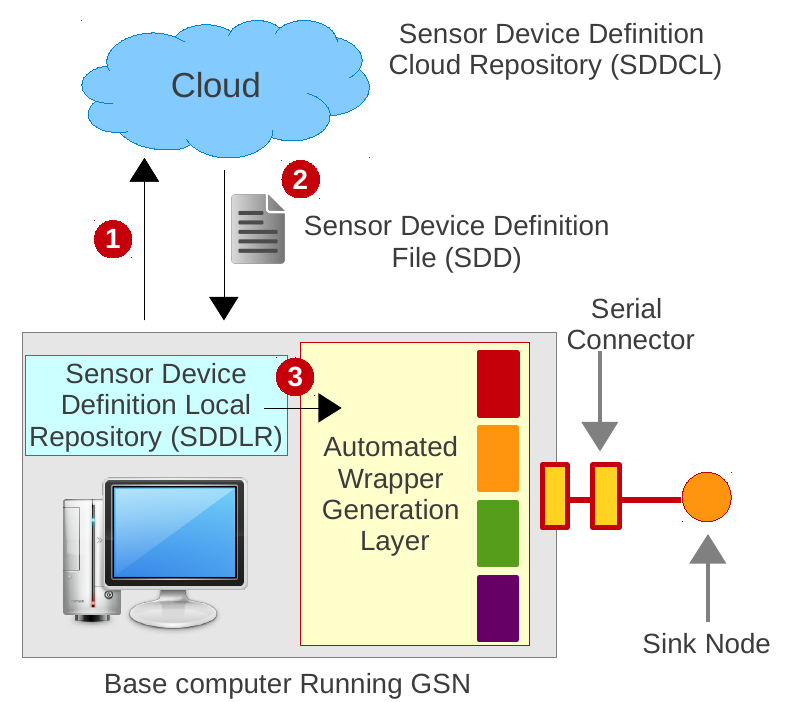}
 \caption{Proposed ASCM4GSN Architecture}
 \label{Extended GSN Data Acquisition Architecture}	
\end{figure}

If Sensor Device Definition Local Repository (SDDLR) does not have a matching SDD file, GSN will automatically connect to the Sensor Device Definition Cloud Repository (SDDCR) and search for a matching wrapper definition file. If there is a matching SDD file, the GSN will fetch the SDD file and feed it to the \textit{Automated Wrapper Generation Layer}. This layer generates the wrapper class, compiles it, and pushes it to the wrapper repository. After that, the wrapper life cycle proceeds as explained in Section \ref{sec:Gsn Wrapper'S Life Cycle}. Figure \ref{Extended GSN Data Acquisition Architecture} shows our proposed architecture.

If there isn't any SDD file in the Sensor Device Definition Cloud Repository (SDDCR) for a specific sensor, then the developers may need to develop a SDD file based on the SDD specification. However, the developers can upload their SDD file to the cloud so other users do not need to develop it again. This approach saves time and cost. In the future, there will be increasing number of sensors available in the market. Our community based cloud approach would be ideal to deal with wider adaptation of IoT and sensor network deployments.

\begin{figure*}[t]
 \centering
 \includegraphics[scale=0.97]{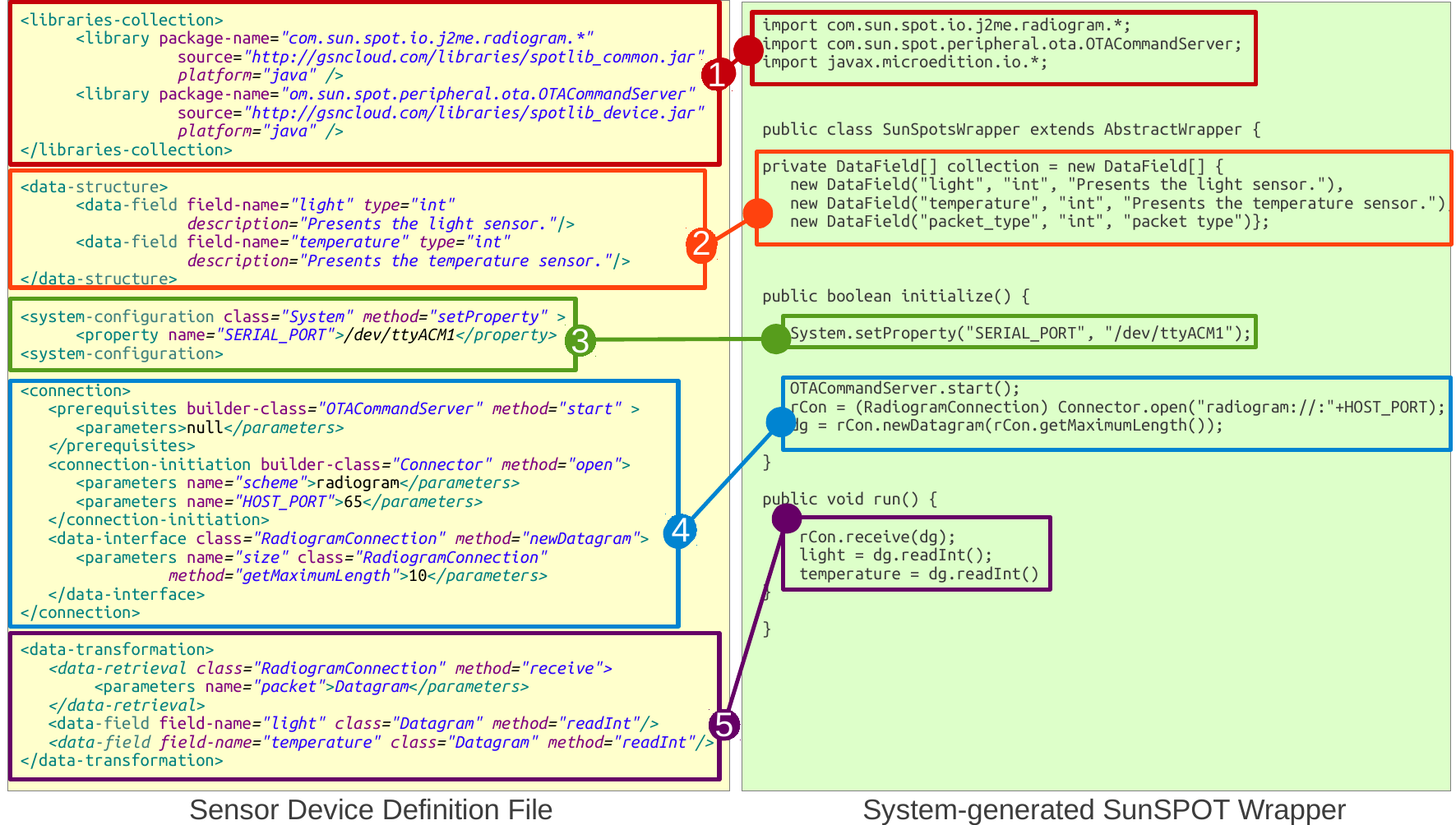}
 \caption{Comparison of SDD File and System-generated Wrapper}
 \label{Comparison of SDD File and System-generated Wrapper}
\vspace{-0.2cm}	
\end{figure*}

There are many reasons to base our approach on SDD files rather than wrapper classes. We could also use direct wrapper classes instead of using SDD files. The reason for adding extra step to our approach can be explained as follows.

If we use a wrapper class, then it would be a platform specific such as Java, C\#, etc. We keep SDD files platform independent or cross platform. Some sensor devices may need to be configured using machine specific details such as port numbers, IP address, etc. Editing a SDD file is a much cleaner and easier way compared to editing a class file. In future, software tools can be developed to make the editing significantly easier. In that case, designing tools to edit platform independent file is much easier than editing different class files written in different programming languages. 

At a later stage it may be required to combine SDD files with semantic technologies in order to make the automation more sophisticate. As we mentioned earlier, wrapper class also need to use third-party libraries which are developed by sensor device manufacturers (e.g. camera). If we based our approach on wrapper classes, then we may need to distribute the libraries with the class file as well. This could create licensing and legal issues. In SDD files, we only need provide the link to the libraries so IoT middleware can download the libraries from the sensor device manufacturers.

%

The second problem we identified has been addressed by connecting GSN to the cloud. We developed a Sensor Device Definition Cloud Repository (SDDCR), which acts as a global repository for SDD files, so software developers and hardware manufacturers can upload SDD files to the cloud. This means if someone develops a SDD file for SunSPOT once, any GSN instance deployed around the world would be able connect SunSPOT sensors automatically using that SDD file.


\section{ASCM4GSN Sensor Device Definition}
\label{sec:Sensor Device Definition}

Sensor Device Definition (SDD) is an XML file which comprises number of elements. Describing each and every sections and elements of SDD is beyond the scope of this paper. Here, we describe how SDD works using a real world example. The left side of Figure \ref{Comparison of SDD File and System-generated Wrapper} shows the SDD file which corresponds to SunSPOT sensor device. The right side of the figure shows the SunSPOT wrapper generated by the \textit{Automated Wrapper Generation Layer} based on the SDD file on the left. Please note that both files are used for demonstration purposes and only contain major sections and elements.

Even though we do not intend to explain each and every element in the two files, a high level explanation can be made as follows. In Section \ref{sec:The Problem}, we identified six major steps that are common across all the sensor platforms. After analysing the steps, we encapsulate them into five segments in the SDD file. The segments are marked 1 to 5 in Figure \ref{Comparison of SDD File and System-generated Wrapper}. The segments are \textit{libraries collection, data structure, system configuration, connection,} and \textit{data transformation}.

The segment (1) contains all the libraries that need to be downloaded for GSN in order to compile the wrapper. It consists of package names and the sources where the files can be downloaded. This section can also be extended to add installation files. For example, if a specific sensor needs to install drivers on the GSN server, this section can provide the source link to download and install the driver automatically. It also consists of platform information. As we are intended to make these SDD files platform independent, the parameters can specify which libraries are related to which platform (e.g. Java, .Net, C, etc.).

The next segment (2) includes the information about the data structures. It can be used to provide all the information required by the GSN in order to create GSN data model. The sensor data will be stored in the data structure created in this segment. The segment (3) is dedicated to store system level configurations. There are configuration settings that need to be configured. The properties and values need be changed depending on the operating system. For example, serial ports are named as /dev/ttyACM in Linux and as COM in Windows.

The connection segment (4) comprises the information related four steps: initiate connection with the sensor devices, initiate data retrieval mechanisms, retrieve sensor data, and close connection. This segment would be a lengthy section. According to our preliminary investigation, most sensors do have these four steps. Final segment (5) is for the data transformation. The retrieved data packets need to be examined and extract the values from them. These values need to be stored in the data structure defined in the segment (2).

If we consider the length and the complexity of the two files, it is true that SDD file is more lengthy and complex. However, we can easily develop a graphical user interface to produce SDD file very easily which will reduce and hide the complexity in major way.

\section{Android To GSN Connectivity Example}
\label{sec:Implementation}

\begin{figure}[h]
 \centering
 \includegraphics[scale=.45]{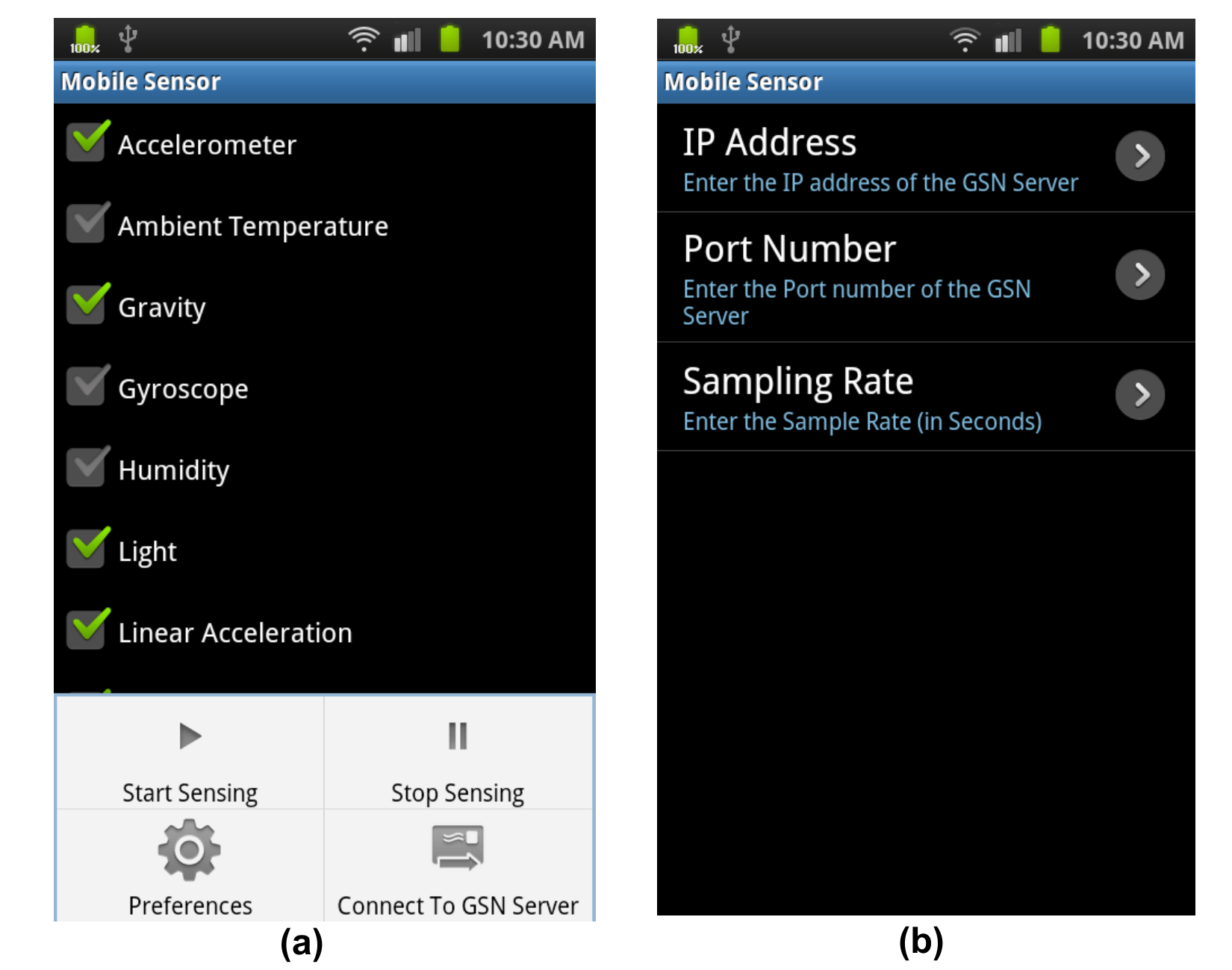}
 \caption{Sensor Data Generator Application}
 \label{Client Application}
\vspace{-0.23cm}	
\end{figure}

We evaluated the process of connecting sensors to an IoT middleware called GSN. \textit{AndroidWrapper} was developed in order to retrieve sensor data from Android mobile phones. It was realised that each and every sensor should have a wrapper talking to a GSN server in order to collect data. Developing such wrappers is a time consuming and tedious job. It could take one or more days for a developer to develop a single wrapper for a specific sensor including the time that would take to familiarise with the specific sensor platform. Our ASCM4GSN approach drastically reduces the development time as the developer does not need to familiarise with the sensor platform in order to generate a wrapper using our approach. Figure \ref{Client Application} shows the client application we developed and installed on Android mobile phones. It generates and sends sensor data packets to the \textit{AndroidWrapper} in GSN.

\section{Related Work}
\label{sec:Related Work}


We recognise two initiatives related to our work; SensorML \cite{P256} and IEEE 1451 standards \cite{P257}. SensorML provides functionalities such as describing sensors and sensor systems for inventory management and sensor discovery, describing sensor geolocation of observed values, describing processes and observations, describing performance characteristics, describing processing and analysis of the sensor observations, etc. IEEE 1451 standards are dedicated to standardise the data communication among sensors, actuators, and related devices. In our solution, we are dealing with sensor installation and configuration challenge where both above initiatives do not explicitly address this challenge.



Californium (Cf) CoAP framework \cite{P074} has proposed a \textit{thin server} architecture to solve the problem of connecting heterogeneous devices from different manufacturers with diverse functionalities to the Internet of things. Cf \cite{P074} puts a \textit{thin server} in front of the device to work as a proxy. Thin server only provides a low-level API to the elementary functionality of a device. The client applications or IoT middleware system can communicate with the device via the thin server using a RESTful API. All the functionalities are encoded as REST resources.

However in this type of approach, the communication between the thin server and the device needs be developed by the developers. This decrease the efficiency in connecting thing to the IoT. Furthermore, using a thin server could also raise issues in term of energy efficiency and communication overheads, specially due to the magnitude of IoT. This approach will perform very well, if the device manufacturer provides a RESTful API integrated to the device's software package itself. Further, this approach would be ideal for device manufacturers to be followed as a step forward in standardising the communication between devices and applications.


Web Services Gateways \cite{P212} is an approach based on Model Driven Architecture (MDA) and Device Profile for Web Services (DPWS). The focus is on connecting industrial devices, where the devices have a lifetime of often more than 40 years, to the client applications in IoT paradigm. They have developed gateways comprises with web services that provide interfaces to access the devices.The web service are generated automatically using predefined models.

In \cite{P212}, performance evaluation results has shown that the approach is not scalable. The papers \cite{P212, P380} admit that the web services approach would perform less compared to native code approach. That is why we proposed a native code generation approach over other mechanisms. The native code approach increases the node complexity and the web service approach limits the scalability. There is a trade of between the two.



InterX \cite{P251} is a smart phone-based service interoperability gateway for heterogeneous smart objects. It employs a mediator gateway that can transform  one protocol to another. For example, InterX enables the communication between Bluetooth based smart object and UPnP based smart object via a gateway in runtime. 

In contrast, our approach is based on development time mediator. We use XML as the mediator to produce the native code that can establish the communication between IoT middleware systems and smart devices. Runtime mediators give less performance due to communication overheads. Another difference is that InterX is focused on well recognised appliances such as digital camera, tv, video player, etc. In our approach, we focused on low level sensors such a SunSPOT, Arduio, and we also offer extensibility to facilitate more low level sensors.


Hydra \cite{P042} is an IoT middleware that allows developers to incorporate heterogeneous physical devices into their applications. The interaction between devices and the middleware are enabled though web services. Hydra is based on a semantic Model Driven Architecture for easy programming. This is similar to the Web Services Gateways \cite{P212} approach we presented earlier. Even though the performance evaluation of the Hydra middleware is not available in device connection perspective, the paper \cite{P212} has raised the similar issues related to employing web services as the gateways for smart devices.


uMiddle \cite{P358} is a bridging framework that enables seamless device interaction over diverse middleware platforms. This approach is similar to the InterX \cite{P251}. uMiddle transforms one protocol to another in runtime. This middleware is focused on the interoperability between popular protocols such as  Bluetooth, UPnP, etc. In contrast, we are more concerned about connecting low level sensors to IoT middleware solutions. uMiddle have identified three essential requirements of an interoperability middleware platform; transport-level bridging, service-level bridging, and device-level bridging. We have also considered these requirements in our approach.


\section{Conclusion and Future Work}
\label{sec:Conclusion}

In this article, we propose and discuss the ASCM4GSN architecture and develop a specification that can be used to automate the sensor data acquisition and configuration process related to IoT middleware. We conducted our implementation and experiments based on a popular IoT middleware platform called GSN. Sensor devices come with APIs that need to be used in order to retrieve sensor data from the sensor devices. For this, middleware has to be aware of sensor devices. Typically, sensor drivers/wrappers need to be developed manually by a programmer using the third-party libraries. This requires more time, cost and effort. We have demonstrated that automating the process of developing sensor drivers/wrappers will  improve efficiency and productivity.  We introduced Sensor Device Description (SDD) specification to facilitate the automation. We have proposed an SDD sharing mechanism using a cloud repository to reduce the repetitive work. Currently, SDD files can be used to generate wrappers for GSN. However,  we intend to keep SDD file as a generic sensor device definition mechanism that would be independent from IoT middleware solutions.

Our future research work aims at efficient and effective automation of connecting things to IoT middleware as well as incorporating generated extended functionality.We will combine context capturing and semantic data technologies with procesing of sensor data inside the wrapper itself.

  \bibliography{Bibliography}
  \bibliographystyle{abbrv}

\end{document}